\documentclass[aps, superscriptaddress]{revtex4-1}
\pdfoutput=1
\usepackage{graphicx}

\usepackage{amsmath}
\usepackage{amssymb}
\usepackage{bm}
\usepackage{color}
\usepackage{comment}

\newcommand{\mean}[1]{\langle{#1}\rangle}

\newcommand{\bra}[1]{\langle{#1}|}
\newcommand{\ket}[1]{|{#1}\rangle}

\newcommand{\Tr}{{\rm Tr}\hspace{0.07cm}}

\newcommand{\abs}[1]{{|#1|}}

\begin{document}
	\title{Quantum asymptotic phases reveal signatures of\\ quantum synchronization}
	\author{Yuzuru Kato}
	\email{Corresponding author: katoyuzu@fun.ac.jp}
	\affiliation{Department of Complex and Intelligent Systems, Future University Hakodate, Hokkaido 041-8655, Japan }
	\author{Hiroya Nakao}
	\affiliation{Department of Systems and Control Engineering,
		Tokyo Institute of Technology, Tokyo 152-8552, Japan}
	\date{\today}

\begin{abstract}
Synchronization of quantum nonlinear oscillators has attracted much attention recently.
To characterize the quantum oscillatory dynamics, we recently proposed a fully quantum-mechanical  definition of the asymptotic phase, which is a key quantity in the synchronization analysis of classical nonlinear oscillators \cite{kato2022definition}. In this work, we further extend this theory and introduce multiple asymptotic phases using the eigenoperators of the adjoint Liouville superoperator of the quantum nonlinear oscillator associated with different fundamental frequencies. We analyze a quantum van der Pol oscillator with Kerr effect in the strong quantum regime and show that the system has several different fundamental frequencies. By introducing order parameters and power spectra in terms of the associated quantum asymptotic phases, we reveal that phase locking of the system with a harmonic drive at several different frequencies, an explicit quantum signature observed only in the strong quantum regime, can be interpreted as synchronization on a torus rather than a simple limit cycle.
\end{abstract}

\maketitle


\section{Introduction}
Spontaneous rhythmic oscillations and synchronization are phenomena ubiquitously observed in various fields of science and technology~\cite{winfree2001geometry, kuramoto1984chemical,  pikovsky2001synchronization, nakao2016phase, ermentrout2010mathematical, strogatz1994nonlinear}.
Owing to the recent progress in nanotechnology,   
synchronization in micro- and nano-scale devices has been realized experimentally \cite{matheny2019exotic, kreinberg2019mutual, singh2019mutual, colombano2019synchronization, sheng2020self}
and experimental demonstrations of quantum phase synchronization in spin-$1$ atoms \cite{laskar2020observation}, in nuclear spin systems \cite{krithika2022observation},
and on the IBM Q system \cite{koppenhofer2020quantum} have been reported.
A number of theoretical investigations have also  been performed to reveal novel quantum signatures in synchronization
\cite{
	lorch2016genuine,
	lorch2017quantum,
	nigg2018observing,
	mari2013measures,
	weiss2016noise,
	hush2015spin,
	weiss2017quantum,
	jaseem2020generalized,
	lee2013quantum,
	walter2014quantum,
	sonar2018squeezing,
	kato2019semiclassical,
	kato2020semiclassical,
	lee2014entanglement,
	witthaut2017classical,
	roulet2018quantum,
	es2020synchronization,
	kato2021enhancement, 
	kato2021instantaneous,
	li2021quantum, 
	xu2014synchronization,
	mok2020synchronization, 
	roulet2018synchronizing,
	chia2020relaxation, 
	arosh2021quantum, cabot2019quantum,cabot2021metastable, galve2017quantum, eneriz2019degree, 
	solanki2022role, setoyama2022quantum, kato2022definition, walter2015quantum}.

In the strong quantum regime where only a small number of energy states participate in the system dynamics,
the discrete nature of the energy spectrum can give rise to explicit quantum signatures,
such as multiple phase locking at several different frequencies~\cite{lorch2016genuine}
and synchronization blockade~\cite{lorch2017quantum, nigg2018observing}.
In characterizing quantum synchronization,
several 
definitions of the system's oscillation phase
have been proposed
~\cite{lorch2016genuine, mari2013measures, weiss2016noise, hush2015spin, weiss2017quantum, galve2017quantum, jaseem2020generalized, eneriz2019degree, 
	solanki2022role, setoyama2022quantum}. 

In classical mechanics,  nonlinear dissipative systems exhibiting spontaneous rhythmic dynamics can be modeled as limit-cycle oscillators.
The asymptotic phase~\cite{winfree2001geometry, kuramoto1984chemical, pikovsky2001synchronization, nakao2016phase, ermentrout2010mathematical}
is a fundamental quantity for the analysis of synchronization, which is defined by the oscillator's vector field and increasing with a constant frequency in the basin of the limit-cycle attractor.
It provides the basis for \textit{phase reduction theory}~\cite{winfree2001geometry, kuramoto1984chemical, pikovsky2001synchronization, nakao2016phase, ermentrout2010mathematical, strogatz1994nonlinear}, 
a standard dimensionality-reduction method to derive phase equations describing weakly-coupled oscillators.

We recently introduced the asymptotic phase for quantum oscillatory systems \cite{kato2022definition} by extending the definition of the asymptotic phase for classical stochastic oscillatory systems \cite{thomas2014asymptotic} based on the Koopman operator theory \cite{kato2021asymptotic}.
This quantum asymptotic phase is defined fully quantum-mechanically in terms of the eigenoperator of the system's adjoint Liouville superoperator associated with the slowest decaying mode
and hence applicable even in the strong quantum regimes where 
we cannot rely on the 
limit-cycle trajectory in the classical limit as in the semiclassical regime~\cite{hamerly2015optical, kato2019semiclassical, kato2020semiclassical}.

In this study, we further extend
this definition and introduce multiple
quantum asymptotic phases in terms of the eigenoperators 
with several different fundamental frequencies,
i.e., with the eigenvalues possessing the smallest absolute imaginary part in the individual branches of the eigenvalues near the imaginary axis,
which play dominant roles in synchronization dynamics.
As an example, we analyze a quantum van der Pol oscillator with Kerr effect and show that it possesses several dominant eigenvalues with different fundamental frequencies in the strong quantum regime.  
By introducing the order parameters and power spectra in terms of the associated quantum asymptotic phases with respective fundamental frequencies, we reveal a torus-like structure of multiple-frequency phase locking of the system with a harmonic drive~\cite{lorch2016genuine},
which is observed only in the strong quantum regime.


\section{Asymptotic phases for quantum oscillatory systems}
In Ref.~\cite{kato2022definition}, we proposed a definition of the asymptotic phase for quantum oscillatory systems in terms of the eigenoperator of the adjoint Liouville superoperator associated with the slowest decaying mode, 
inspired by the definition of the asymptotic phase for classical stochastic oscillators 
~\cite{thomas2014asymptotic, kato2021asymptotic}.
In this section, we briefly review this definition and extend it to introduce multiple quantum asymptotic phases by using the eigenoperators 
associated with several eigenvalues with different fundamental frequencies.

\subsection{Quantum master equation} 
We consider a quantum nonlinear oscillatory system with a single degree of freedom coupled to reservoirs,
and assume that the interactions of the system with the reservoirs are instantaneous and Markovian approximation can be employed.
The time evolution of the system's density operator $\rho$ is described by a quantum master equation~\cite{carmichael2007statistical, gardiner1991quantum},
\begin{align}
\label{eq:me}
\dot{\rho}
= \mathcal{L}\rho =
-i[H, \rho] 
+ \sum_{j=1}^{n} \mathcal{D}[C_{j}]\rho,
\end{align}
where $\mathcal{L}$ is a Liouville superoperator representing 
the evolution of $\rho$,
$H$ is a system Hamiltonian, 
$C_{j}$ is a coupling operator between the system  and 
$j$th reservoir $(j=1,\ldots,n)$,  
$[A,B] = AB-BA$ is the commutator,
$\mathcal{D}[C]\rho = C \rho C^{\dag} - (\rho C^{\dag} C + C^{\dag} C \rho)/2$ 
is the Lindblad form ($\dag$ denotes Hermitian conjugate), 
and the reduced Planck's constant is set as $\hbar = 1$.

Introducing an inner product $\mean{X,Y}_{tr} =\Tr{ ( X^{\dag}Y )}$
of linear operators $X$ and $Y$, we  define the adjoint superoperator $\mathcal{L}^{*}$ of $\mathcal{L}$ satisfying
$\mean{\mathcal{L}^{*}X,Y}_{tr}
=\mean{X,\mathcal{L}Y}_{tr}$
\begin{align}
\label{eq:bme}
\mathcal{L}^{*} X = i[H, X] + \sum_{j=1}^{n} \mathcal{D}^{*}[C_{j}]X,
\end{align}
where $\mathcal{D}^{*}[C]X = C^{\dag} X C - (X C^{\dag} C + C^{\dag} C X)/2$ is the adjoint Lindblad form.
This ${\cal L}^{*}$ describes the evolution of an observable $F$ as 
\begin{align}
\dot{F} = {\cal L}^{*} F.
\label{eq:bme2}
\end{align}
In the Schr\"odinger picture, the density operator $\rho$ evolves as in Eq.~(\ref{eq:me}) while the observation operator $F$ does not vary with time, and in the Heisenberg picture, $F$ evolves as in Eq.~(\ref{eq:bme2}) while $\rho$ remains constant.
The expectation value 
\begin{align}
\langle F \rangle = \mbox{Tr}(\rho F) = \mean{\rho, F}_{tr}
\end{align}
of $F$ with respect to $\rho$ is kept the same in both pictures
(note that $\rho$ is self-adjoint).

We assume that the Liouville superoperator $\mathcal{L}$ has a set of eigensystem (an eigenvalue and right and left eigenoperators) $\{ \lambda_{k}, U_{k}, V_{k} \}$ 
satisfying
\begin{align}
\mathcal{L} U_{k} =  \lambda_{k} U_{k},
\quad
\mathcal{L}^{*} V_{k} =  \overline{\lambda_{k}} V_{k},
\quad
\mean{V_{k}, U_{l}}_{tr} = \delta_{kl},
\label{eq:eigentriplet1}
\end{align}
for $k, l=0, 1, 2, \ldots$, where the overline indicates complex conjugate~\cite{li2014perturbative}.
We assume that among $\{ \lambda_k \}_{k \geq 0}$, one eigenvalue $\lambda_0$ is always $0$, which corresponds to the stationary state $\rho_0$ of the system satisfying ${\cal L} \rho_0 = 0$, 
and all other eigenvalues have negative real parts.
This assumption also means that the system has no decoherence-free subspace \cite{lidar1998decoherence}. 
Considering the system's oscillatory dynamics, we also assume that the eigenvalues of ${\cal L}$ 
with the largest non-vanishing real part (i.e., with the slowest decay rate) are given by a complex-conjugate pair and denote them as $\Lambda_1$ and $\overline{\Lambda_1}$, where $\Omega_1 = \mbox{Im}\ \overline{\Lambda_1}$ gives the frequency of the slowest decay mode.
One may also choose $\Omega_1 = \mbox{Im}\ \Lambda_1$, which reverses the direction of the asymptotic phase.
The sign of $\Omega_1$ can be chosen arbitrarily and will be fixed later.

Figure~\ref{fig_1} shows typical examples of the eigenvalues of ${\cal L}$ of the quantum vdP oscillator (see the next section for details). 
The eigenvalues form several branches.
In the semiclassical regime (Fig.~\ref{fig_1}(b)), the rightmost branch is far apart from the other branches and thus it is the only dominant branch, while in the strong quantum regime (Fig.~\ref{fig_1}(a)), the branches are closer to each other and the system possess several comparably important branches.
It can also be seen in the semiclassical regime  (Fig.~\ref{fig_1}(b)) that
the imaginary part of the eigenvalues are approximately integer multiples of the fundamental frequency (the smallest absolute imaginary part with the slowest decay rate) in each individual branch.

\subsection{Phase-space representation}

The density operator $\rho$ can also be transformed into a quasiprobability distribution in the phase space~\cite{carmichael2007statistical, gardiner1991quantum, cahill1969density}.
We use the $P$-representation and  describe $\rho$ as
\begin{align}
\rho = \int p({\bm \alpha}) | \alpha \rangle \langle \alpha |  d {\bm \alpha},
\end{align}
where $| \alpha \rangle$ is a coherent state specified by 
a complex value $\alpha$, or equivalently by a complex vector $\bm{\alpha} = (\alpha, \overline{\alpha})^{T}$,
$p({\bm \alpha})$ is a quasiprobability 
distribution of ${\bm \alpha}$, $d{\bm \alpha} = d\alpha d\overline{\alpha}$, 
and the integral is taken over the entire complex plane.
The observable $F$ is also transformed into a function in the phase space as
\begin{align}
f({\bm \alpha}) = \langle \alpha | F | \alpha \rangle,
\end{align}
where the operator $F$ is arranged in the normal order~\cite{carmichael2007statistical, gardiner1991quantum, cahill1969density}.
By introducing the $L^2$ inner product $\mean{g(\bm{\alpha}), h(\bm{\alpha})}_{\bm{\alpha}} = \int \overline{ g(\bm{\alpha}) } h(\bm{\alpha}) d \bm{\alpha}$
of two functions $g(\bm{\alpha})$ and $h(\bm{\alpha})$, the expectation value of $F$ with respect to $\rho$ is expressed as
\begin{align}
\langle F \rangle = \mbox{Tr} (\rho F) = \int d{\bm \alpha} p({\bm \alpha}) f({\bm \alpha}) = \mean{p(\bm{\alpha}), f(\bm{\alpha})}_{\bm{\alpha}}.
\end{align}

The time evolution of $p({\bm \alpha})$ corresponding to Eq.~(\ref{eq:me}) is described by a partial differential equation
\begin{align}
\partial_t{p}({\bm \alpha}) = L_{\bm \alpha} p({\bm \alpha}),
\label{eq:qpde}
\end{align}
where the differential operator ${L}_{\bm \alpha}$ 
satisfies $\mathcal{L}\rho = \int {L}_{\bm \alpha} p({\bm \alpha}) | \alpha \rangle \langle \alpha | d{\bm \alpha}$.
The explicit form of ${L}_{\bm \alpha}$ can be calculated from Eq.~(\ref{eq:me})
by using the standard calculus for the phase-space representation
\cite{carmichael2007statistical, gardiner1991quantum, cahill1969density}.
The corresponding evolution of $f({\bm \alpha})$ in the Heisenberg picture is given by
\begin{align}
\partial_t f({\bm \alpha}) = {L}^{+}_{\bm \alpha} f({\bm \alpha}),
\end{align}
where the differential operator ${L}^{+}_{\bm \alpha}$ is 
the adjoint of ${L}_{\bm \alpha}$
with respect to the $L^2$ inner product, i.e., 
$ \mean{ {L}^{+}_{\bm \alpha} g(\bm{\alpha}), h(\bm{\alpha})}_{\bm{\alpha}}
=\mean {g(\bm{\alpha}), {L}_{\bm \alpha} h(\bm{\alpha})}_{\bm{\alpha}}$,
which satisfies ${L}^{+}_{\bm \alpha} f({\bm \alpha}) = 
 \langle \alpha | \mathcal{L}^{*} F | \alpha \rangle$.

The differential operator ${L}_{\bm \alpha}$ also has a set of eigensystem (an eigenvalue and right and left eigenfunctions) $\{ \lambda_{k}, {u}_{k}({\bm \alpha}), {v}_{k}({\bm \alpha}) \}$ which satisfies
\begin{align}
&{L}_{\bm \alpha} {u}_{k} = \lambda_{k} {u}_{k},
\quad
{L}^{+}_{\bm \alpha} {v}_{k} = \overline{\lambda_{k}} {v}_{k},
\quad
\mean{ {v}_{k}, {u}_{l}}_{\bm \alpha} 
= \delta_{kl}.
\end{align}
This eigensystem has a one-to-one correspondence with Eq.~(\ref{eq:eigentriplet1}), where
the eigenvalues $\{ \lambda_{k} \}_{k \geq 0}$ are the same as those of ${\cal L}$; the eigenfunctions ${u}_{k}$ and ${v}_{k}$ of ${L}_{\bm \alpha}$ and ${L}^{+}_{\bm \alpha}$ are related to
the eigenoperators $U_{k}$ and $V_{k}$ of $\mathcal{L}$ and $\mathcal{L}^{*} $ as
\begin{align}
U_{k} = 
 \int {u}_{k}({\bm \alpha}) 
| \alpha \rangle \langle \alpha |
d{\bm \alpha},
\quad
{v}_{k}({\bm \alpha}) = 
\langle \alpha | V_{k} | \alpha \rangle,
\end{align}
which follow from
$\mathcal{L} U_{k}
= 
\int {u}_{k}({\bm \alpha}) \left\{ \mathcal{L} 
| \alpha \rangle \langle \alpha | \right\} d{\bm \alpha}
= 
\int \left\{ {L}_{\bm \alpha} {u}_{k}({\bm \alpha}) \right\}
| \alpha \rangle \langle \alpha | d{\bm \alpha}
=
\lambda_{k} U_{k}
$
and
$
{L}^{+}_{\bm \alpha} {v}_{k} = {L}^{+}_{\bm \alpha} \langle \alpha | V_{k} | \alpha \rangle
= 
\langle \alpha | {\mathcal L}^* V_{k} | \alpha \rangle = \overline{\lambda_{k}} \langle \alpha | V_{k} | \alpha \rangle
=
\overline{\lambda_{k}} {v}_{k}.
$

\subsection{Quantum asymptotic phase associated with the slowest decaying mode}

In Ref.~\cite{kato2022definition}, we defined the quantum asymptotic phase function $\Phi_1$ 
of the system state $\rho$ as the argument (polar angle) of the
expectation of the eigenoperator ${V}_1$ associated with the 
eigenvalue $\Lambda_1$ satisfying ${\cal L}^* {V}_1 =  \overline{\Lambda_1}  {V}_1$, namely,
\begin{align}
	\label{eq:qiso}
\Phi_1(\rho) = \arg \langle V_1 \rangle = \arg \langle \rho, V_1 \rangle_{tr} = \arg \langle p({\bm \alpha}), v_1({\bm \alpha}) \rangle_{\bm \alpha}.
\end{align}
We showed that this quantum asymptotic phase yields appropriate phase values, namely, it always increases with a constant frequency $\Omega_1$ as $\dot{\Phi}_1(\rho) = \Omega_1$ 
with the evolution of $\rho$ even in the strong quantum regime and reproduces the conventional asymptotic phase in the semiclassical regime~\cite{kato2022definition}.
We note that $\Phi_1(\rho)$ is not defined when $\langle V_1 \rangle = 0$.

This quantum asymptotic phase is a natural extension of the asymptotic phase for stochastic limit-cycle oscillators defined in terms of the slowest decaying eigenfunction of the backward Kolmogorov (Fokker-Planck) operator \cite{thomas2014asymptotic} from the Koopman operator viewpoint \cite{kato2021asymptotic}.

\subsection{Multiple quantum asymptotic phases associated with different fundamental frequencies}

In this study, we further extend the definition in the previous subsection and introduce multiple quantum asymptotic phases associated with several different fundamental frequencies by using the principal eigenvalues on different eigenvalue branches near the imaginary axis,
and use them to characterize quantum signatures of synchronization in the strong quantum regime.

As shown in Fig.~\ref{fig_1}(a), in the strong quantum regime, multiple branches of the eigenvalues with different fundamental frequencies can exist near the imaginary axis, suggesting that not only the eigenvalue $\Lambda_1$ on the rightmost branch but also the eigenvalues with the slowest decay rates on the other branches play important roles; for comparison, see Fig.~\ref{fig_1}(b) for a typical example of the eigenvalue $\Lambda_1$ in the semiclassical regime, where only a single dominant branch of eigenvalues exist.
We denote these eigenvalues by 
$\Lambda_1, \Lambda_2, \Lambda_3, \ldots$
and their imaginary parts, i.e., fundamental frequencies, by 
$\Omega_j = \mbox{Im}\ \overline{ \Lambda_j }$ ($j \geq 1$), and call $\{ \Lambda_{j} \}_{j \geq 1}$ the \textit{principal eigenvalues}. 
Here, the first principal eigenvalue $\Lambda_1$ and the fundamental frequency $\Omega_1$ are 
those introduced in the previous subsection.
These principal eigenvalues on the individual branches are shown by red dots in Fig.~\ref{fig_1}(a).

In a similar manner to the phase function $\Phi_1$,
we introduce the $j$-th quantum asymptotic phase function $\Phi_j$ ($j=2, 3, \ldots$) 
of $\rho$ as the argument of the $P$-representation of the eigenoperator ${V}_j$ associated with the principal eigenvalue $\Lambda_j$ satisfying $
{\cal L}^* {V}_j =  \overline{\Lambda_j}  {V}_j$, namely,
\begin{align}
	\label{eq:qiso2}
\Phi_j(\rho) = \arg \langle V_j \rangle = \arg \langle \rho, V_j \rangle_{tr} = \arg \langle p({\bm \alpha}), v_j({\bm \alpha}) \rangle_{\bm \alpha}.
\end{align}
Note that $\Phi_j(\rho)$ is not defined when $\langle V_j \rangle = 0$.
Since
\begin{align}
\frac{d}{dt} \langle V_j \rangle 
= \langle \dot\rho, V_j \rangle_{tr} 
= \langle {\cal L} \rho, V_j \rangle_{tr} 
= \langle \rho, {\cal L}^* V_j \rangle_{tr} 
= \overline{\Lambda_j} \langle \rho, V_j \rangle_{tr}
= \overline{\Lambda_j} \langle V_j \rangle,
\end{align}
we obtain $\langle V_j \rangle(t) = \langle V_j \rangle(0) e^{\overline{\Lambda_j} t}$ and therefore
\begin{align}
\frac{d}{dt} \Phi_j(\rho) = \frac{d}{dt} \arg \langle V_j \rangle = \mbox{Im}\ \overline{\Lambda_j} = \Omega_ j.
\end{align}
Thus, $\Phi_j$ always increases {with} a constant frequency $\Omega_j$ with the evolution of $\rho$ and plays the role of the asymptotic phase for any $j=1, 2, 3, \ldots$. 

It should be noted that  $\Phi_j(\rho)$ cannot be defined for the stationary state $\rho_0$, because $\rho_0$ is the eigenfunction of the Liouville superoperator $\mathcal{L}$ with the eigenvalue $\lambda_0 = 0$ and hence $\langle V_j \rangle = \langle \rho_0, V_j \rangle_{tr} = 0$ from the biorthogonality in Eq.~(\ref{eq:eigentriplet1}).
 As will be shown later, for the stationary state of a periodically driven system, $\langle V_j \rangle$ takes a non-zero value and  the above definition of the phase functions can be used for the analysis of phase locking.

We stress that in the classical deterministic limit, $\Phi_{j}$ ($j \geq 2$) is not independent from $\Phi_1$ and does not provide additional information, because $\Omega_j$ ($j \neq 1$) is equal to $\Omega_1$ according to the Koopman operator theory
\cite{mauroy2013isostables, mauroy2020koopman, shirasaka2017phase, kuramoto2019concept}.
In the semiclassical regime, $\Omega_j$ ($j \neq 1$) differs only slightly from $\Omega_1$ as shown in Fig.~\ref{fig_1}(b) due to the effect of weak quantum noise and the corresponding decay rate for $j \geq 2$ is much larger, so the corresponding mode does not play an important role.
However, as we see in the next section, $\Phi_{j}$ ($j \geq 2$) is distinctly different from $\Phi_1$ and the corresponding decay rate $\mbox{Re}\ \overline{\Lambda_j}$  ($j \geq 2$) is comparable to $\mbox{Re}\ \overline{\Lambda_1}$ in the strong quantum regime, hence the phase function $\Phi_{j}$ ($j \geq 2$) yields independent information from $\Phi_1$.
%


\section{Quantum van der Pol oscillator with Kerr effect}

\subsection{Eigenvalues of the Liouville superoperator}

As an example, we consider a quantum van der Pol oscillator with Kerr effect.
The master equation is given by~\cite{kato2020semiclassical,lorch2016genuine, kato2022definition}
\begin{align}
\label{eq:qvdp_me}
\dot{\rho} 
= \mathcal{L}_0 \rho,
\quad
\mathcal{L}_0  \rho
= - i \left[ 
H
,\rho\right]
+ \gamma_{1} \mathcal{D}[a^{\dag}]\rho + \gamma_{2}\mathcal{D}[a^{2}]\rho,
\end{align}
where $H = \omega_0 a^{\dag}a + K a^{\dag 2} a^2$, 
$\omega_{0}$ is the natural frequency of the oscillator, 
$K$ is the Kerr parameter,
and $\gamma_{1}$ and $\gamma_{2}$ are the decay rates for 
negative damping and nonlinear damping, respectively.
We added a subscript $0$ to the Liouville operator as we will introduce an additional external drive later.

We first consider a strong quantum regime with large $\gamma_{2}$ and $K$, 
where only a small number of energy states participate in the system dynamics 
and the discrete nature of the energy spectrum plays important roles in the dynamics. 
We set the parameters as $\gamma_1=0.1$ and $(\omega_0, \gamma_{2}, K)/\gamma_{1} = (300, 4, 100)$, which are the same as in our previous study \cite{kato2022definition}.
In the numerical calculation, we approximately truncated the density operator as a large-dimensional $N \times N$ matrix and mapped it into a $N^2$-dimensional vector in the double-ket notation~\cite{albert2018lindbladians}.
We can then represent the Liouville operator by a $N^2 \times N^2$ matrix and obtain 
the asymptotic phase in Eq.(\ref{eq:qiso}) from the eigensystem of this matrix \cite{kato2022definition}. 

\begin{figure} [!t]
	\begin{center}
		\includegraphics[width=1\hsize,clip]{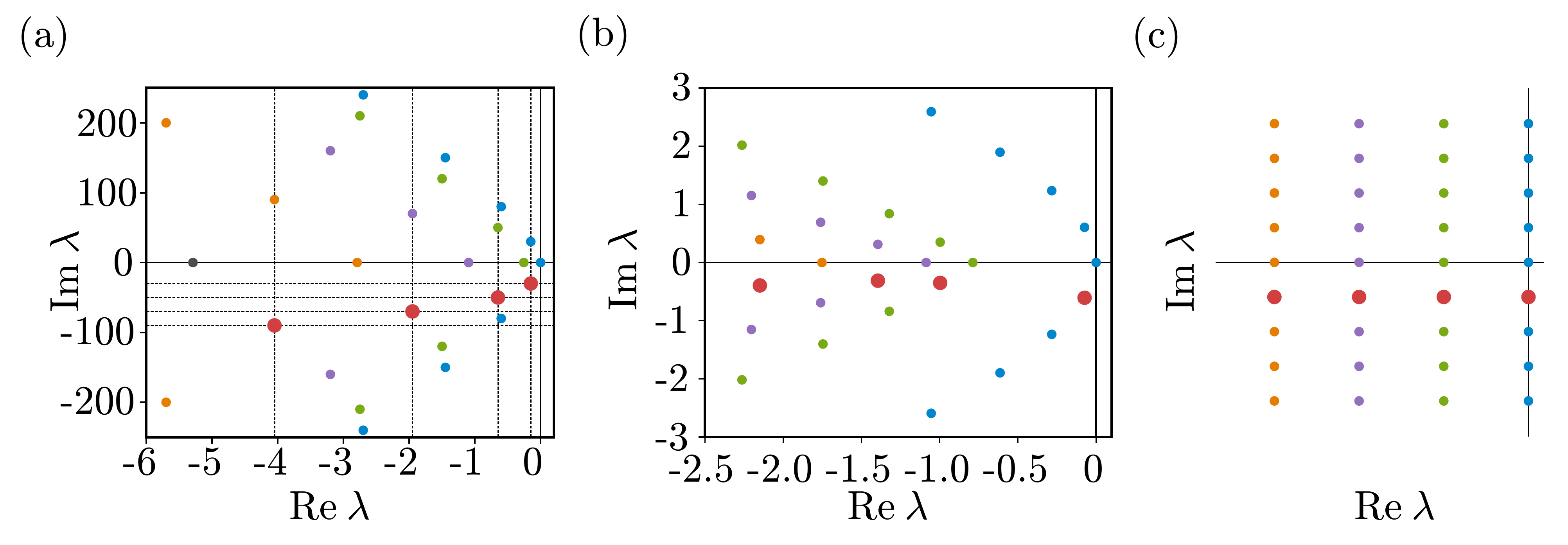}
		\caption{
			Eigenvalues of the Liouville superoperator  $\mathcal{L}_0$ near the imaginary axis. 
			(a) Strong quantum regime.
			(b) Semiclassical regime.
			The red dot on the light-blue branch represents the principal eigenvalue $\Lambda_1$ with the the slowest decay rate. 
			(c) Classical limit (schematic). 
			In (a, b, c),  individual branches of eigenvalues are shown with different colors and 
			the red dots represent the principal eigenvalues $\Lambda_j$ ($j=1,2,3,4$ from the right) with the fundamental frequencies in individual branches.
			In (a), the dotted lines indicate $\widetilde{\lambda}_m$ ($m=0,1,2,3$) in Eq.~(\ref{qvdp_eig}).
			}
		\label{fig_1}
	\end{center}
\end{figure}

Figure~\ref{fig_1}(a) shows the eigenvalues of ${\cal L}_0$ near the imaginary axis obtained numerically.
We can identify several branches of the eigenvalues near the imaginary axis
characterized by the principal eigenvalues $\Lambda_1, \Lambda_2, \Lambda_3, \ldots$
whose fundamental frequencies $\Omega_1 = \mbox{Im}\ 
\overline{\Lambda_1}, \Omega_2 = \mbox{Im}\ \overline{\Lambda_2}, \Omega_3 = \mbox{Im}\ \overline{\Lambda_3}, \ldots$ are different from each other. 
Moreover, their decay rates, which are characterized by $\mbox{Re}\ \overline{\Lambda_1}, \mbox{Re}\ \overline{\Lambda_2}, \mbox{Re}\ \overline{\Lambda_3}, \ldots$ are comparable to each other.
This indicates that not only the quantum asymptotic phase $\Phi_1$ associated with the principle eigenvalue $\Lambda_1$ of the slowest decaying mode~\cite{kato2022definition} but also those associated with the other principal eigenvalues $ \Lambda_2, \Lambda_3, \ldots$ can play important roles in the dynamics.
We choose a negative value for each $\Omega_{j}$ so that the corresponding $\Phi_j$ increases from $0$ to $2\pi$ in the counterclockwise direction.

In Ref.~\cite{lorch2016genuine}, it is shown that, in the strong quantum regime,
$\ket{m+1}\bra{m}$ is an approximate eigenoperator of ${\cal L}_0$ with the eigenvalue
\begin{align}
	\label{qvdp_eig}
	\widetilde{\lambda}_m=i[-\omega_0-2mK]- \frac{1}{2} \{ \gamma_{1}(2 m+3)+2 \gamma_{2} m^{2} \},
\end{align}
where $m=0, 1, 2, \ldots$, 
namely, $\mathcal{L}^{-1}_0  \ket{m+1}\bra{m} \approx \widetilde{\lambda}^{-1}_m  \ket{m+1}\bra{m} $. As shown in Fig.~\ref{fig_1}(a), these eigenvalues correspond to the principal eigenvalues of ${\cal L}_0$, i.e., $\Lambda_{j} \approx \widetilde{\lambda}_{j-1}$, and thus $V_{j} \approx \ket{j}\bra{j-1}$ ($j=1, 2, 3, 4$).
This indicates that that periodic transitions between the adjacent discrete energy states play important roles in the oscillatory behavior in the strong quantum regime, where the difference in the transition frequencies arises due to the unequal spacing of the energy levels characterized by the Kerr parameter in Eq.~(\ref{qvdp_eig}).

As a comparison, we next consider the semiclassical regime where $\gamma_{2}$ and $K$ are sufficiently small 
and the semiclassical approximation can be taken \cite{kato2019semiclassical}.
We set the parameters as $\gamma_1=1$ and$(\omega_0, \gamma_{2}, K)/\gamma_{1} = (0.1, 0.05, 0.025)$, which are the same as those used in Ref.~\cite{kato2022definition}.
In this regime, we can approximate Eq.~(\ref{eq:qpde}) by a quantum Fokker-Planck equation for $p({\bm \alpha})$ and the system can be approximately described as a Stuart-Landau oscillator (Hopf normal form) subjected to small quantum noise \cite{kato2020semiclassical,lorch2016genuine, kato2022definition}. (Therefore, the conventional quantum vdP oscillator is also called a quantum Stuart-Landau oscillator recently~\cite{mok2020synchronization, chia2020relaxation}  and a more appropriate model of the quantum van der Pol oscillator has also been proposed \cite{chia2020relaxation, arosh2021quantum}.)

Figure~\ref{fig_1}(b) shows the eigenvalues of $\mathcal{L}_0$ near the imaginary axis,
where the principal eigenvalues are shown by red dots on the individual branches; $\Lambda_1 = \mu_1 + i \Omega_1$  ($\mu_1 < 0$) is on the rightmost light-blue branch.
In contrast to the strong quantum regime, the rightmost branch of the eigenvalues, approximately given by a parabola $\hat{\lambda}_n = i \Omega_1 n + \mu_1 n^2~ (n=0, \pm 1, \pm 2, \ldots)$ 
passing through $\overline{\Lambda_1}$, is isolated from other branches of eigenvalues with faster decay rates; the relative decay rate $\mbox{Re}~\overline{\Lambda_2}/\mbox{Re}~\overline{\Lambda_1} \approx -0.996/(-0.0736) \approx 13.5$ is more than three times larger than $\mbox{Re}~\overline{\Lambda_2}/\mbox{Re}~\overline{\Lambda_1} \approx -0.65/(-0.15) \approx 4.33$ in the strong quantum regime in Fig.~\ref{fig_1}(a), indicating that only the rightmost branch is dominant in the semiclassical regime.
Also, the fundamental frequencies of the other branches, defined as the smallest absolute imaginary part of the eigenvalues, are approximately equal to $\Omega_1$; the small differences in the fundamental frequencies arise from small quantum noise.
Thus, it is sufficient to consider only $\Lambda_1$ and introduce a single phase function $\Phi_1$ in this regime. 

The system in the classical limit, i.e., in the limit of vanishing quantum noise, 
is described by the drift term of the approximate quantum Fokker-Planck equation for $p({\bm \alpha})$,
which represents the Stuart-Landau oscillator
and possesses a stable limit-cycle solution \cite{kato2020semiclassical,lorch2016genuine, kato2022definition}.
Figure~\ref{fig_1}(c) shows a schematic diagram of the eigenvalues of the differential operator ${L}^{+}_{\bm \alpha}$, which are equivalent to those of the backward Liouville operator {$\mathcal{L}^{*}_0$}, in the classical limit.
They are given in the form 
$\lambda_c = m \kappa_1  + i n \omega_1~ (m=0,1,2,\ldots, n = 0, \pm 1, \pm2)$,
where $\kappa_1$ is the real part of the largest negative eigenvalue 
and $\omega_1$ is the pure-imaginary eigenvalue with the smallest 
absolute imaginary part.
Thus, $\Phi_{j}$ ($j \geq 2$) is identical to $\Phi_{1}$ and does not provide additional information, because $\Omega_j$ ($j \neq 2$) is identical to $\Omega_1$.

\begin{figure} [!t]
	\begin{center}
		\includegraphics[width=0.5\hsize,clip]{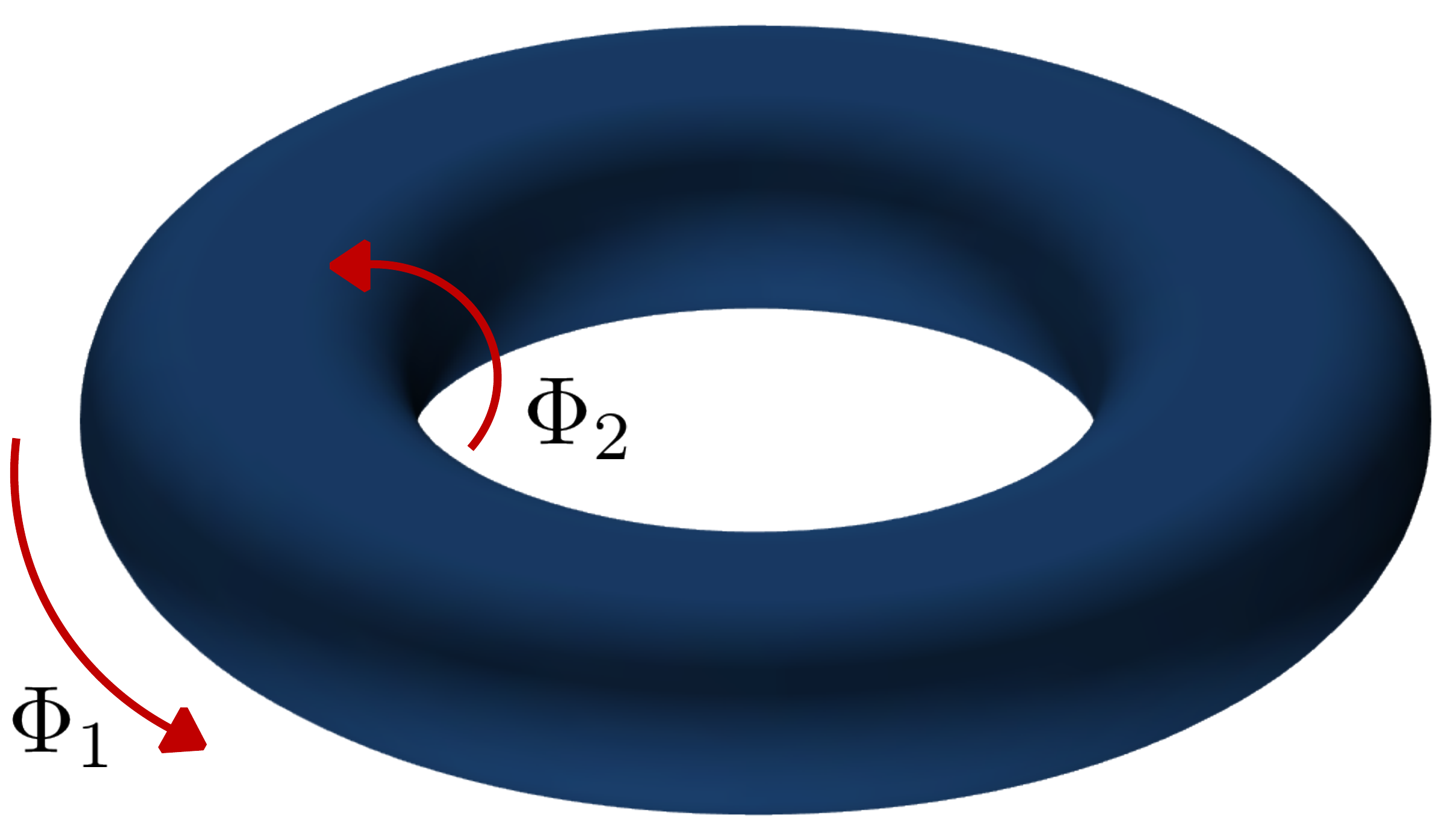}
		\caption{
			Schematic diagram of the torus behavior of the system with two fundamental frequencies.
		}
		\label{fig_2}
	\end{center}
\end{figure}

The above results suggest that the phase function $\Phi_{j}$ yields independent information from $\Phi_1$ only in the strong quantum regime.
The existence of several dominant fundamental frequencies in the strong quantum regime
suggests that the system behaves like a torus rather than a limit cycle with a single fundamental frequency, and that we need to consider the phase functions $\Phi_2, \Phi_3, \ldots$ associated with $\Lambda_2, \Lambda_3, \ldots$ in addition to $\Phi_1$ and $\Lambda_1$.
Figure~\ref{fig_2} shows a schematic picture of the torus behavior of the system with two fundamental frequencies.

\begin{figure} [htbp]
	\begin{center}
		\includegraphics[width=1\hsize,clip]{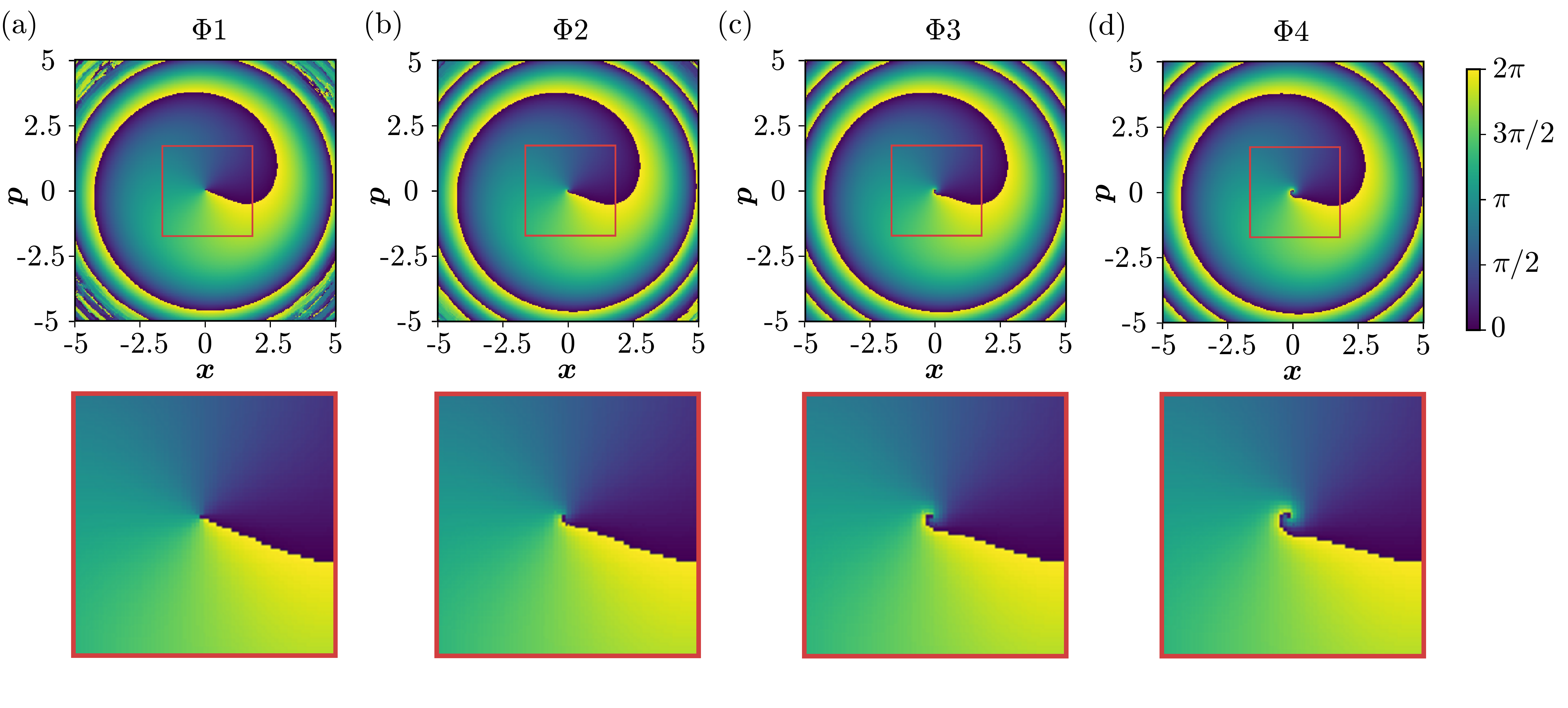}
		\caption{
			Asymptotic phase functions of the quantum van der Pol oscillator with Kerr effect
			in the strong quantum regime. 
			(a) $\Phi_1$, (b) $\Phi_2$, (c) $\Phi_3$, and (d) $\Phi_4$.
			The parameters are $\gamma_1=0.1$ and $(\omega_0, \gamma_{2}, K)/\gamma_{1} = (300, 4, 100)$.
			Figures in the bottom row show enlargements of the regions near the origin in the corresponding figures in the top row.
			In all figures, $(x,p)=(2.5, 0)$ is chosen as the phase origin where $\Phi_j = 0$ ($j=1,2,3,4$).
		}
		\label{fig_3}
	\end{center}
\end{figure}

\subsection{Asymptotic phase functions in the strong quantum regime}

In this subsection, we examine the validity of the quantum asymptotic phase functions in the strong quantum regime.

Figure~\ref{fig_3} shows the asymptotic phase functions 
$\Phi_j({\bm \alpha})$ for $j=1, 2, 3,$ and $4$ of the pure coherent state ${\bm \alpha}$ on the complex plane $(x = \mbox{Re}\ \alpha, p = \mbox{Im}\ \alpha)$.
The parameters are the same as in Fig.~1(a).  
These asymptotic phase functions look similar, but they are associated with different fundamental frequencies and slightly different from each other near the origin as shown in the enlarged figures. Thus, they capture different oscillatory dynamics of the system.
Though not shown, we may also draw similar asymptotic phase functions $\Phi_j$ for $j \geq 5$, which are less dominant.
Here, the asymptotic phase functions for different values of $j$
look similar to each other since they are identical in the classical limit \cite{mauroy2013isostables, mauroy2020koopman, shirasaka2017phase, kuramoto2019concept}, but the differences in their eigenfrequencies brought by the strong quantum effect take an important role in analyzing strong quantum signatures in synchronization.
 
\begin{figure} [!t]
	\begin{center}
		\includegraphics[width=0.9\hsize,clip]{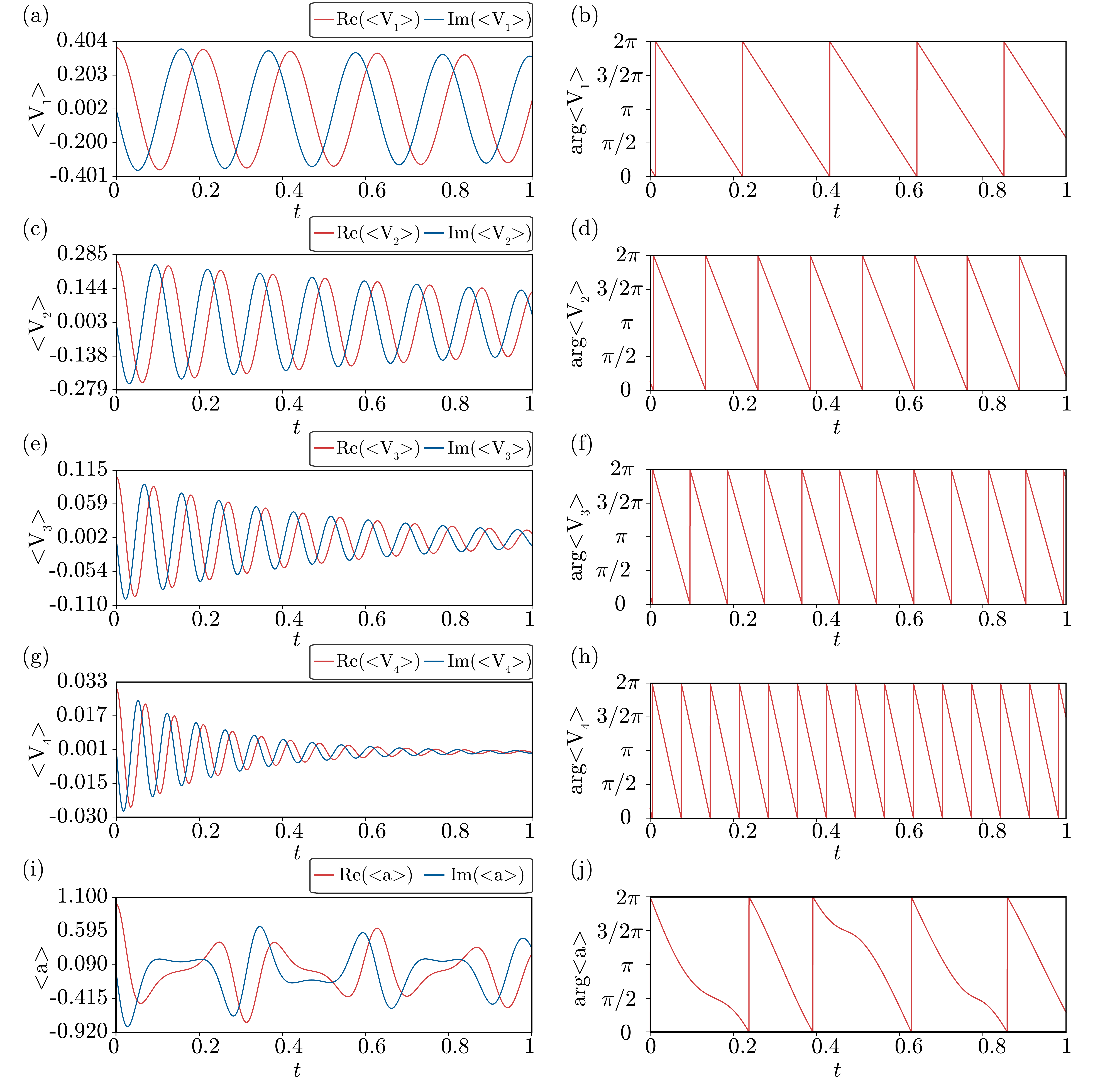}
		\caption{
			Evolution of the expectation values of $V_j$ ($j=1,2,3,4$) and the annihilation operator $a$ (a, c, e, g, i) and their arguments (b, d ,f, h, j) from a pure coherent initial state in the strong quantum regime. 
			The parameters are $\gamma_1=0.1$ and $(\omega_0, \gamma_{2}, K)/\gamma_{1} = (300, 4, 100)$.
			(a, c, e, g) $\mean{V_j}$,  (b, d, f, h) $\arg \mean{V_j}$, (i) $\mean{a}$,  (j) $\arg \mean{a}$. 
		}
		\label{fig_4}
	\end{center}
\end{figure}

To demonstrate that these quantum asymptotic phase functions yield appropriate phase values, we consider free oscillatory relaxation of $\rho$ from a pure coherent initial state $\rho =  \ket{\alpha_0}\bra{\alpha_0}$ with $\alpha_0=1$ at $t=0$ and measured the evolution of the expectation values of $V_j$ and their arguments, i.e., their asymptotic phases $\Phi_j(\rho) = \arg \langle V_j \rangle$ $(j=1,2,3,4)$, as well as those of the annihilation operator $a$ for comparison.
It is noted that, though the initial condition is a pure coherent state, the system state quickly becomes mixed due to the interaction with the reservoirs. 

We can confirm that each $\langle V_j \rangle$ exhibits exponentially damped harmonic oscillations as shown in Fig.~\ref{fig_4}(a, c, e, g), and correspondingly, each $\Phi_j(\rho)$
gives constantly varying phase values with frequency $\Omega_j$ as shown in Fig.~\ref{fig_4}(b, d, f, h), verifying the validity of the definition of the quantum asymptotic phases.
It is noted that different asymptotic phases independently capture different oscillation modes in the evolution of $\rho$. 
In contrast, $\langle a \rangle$ shown in Fig.~\ref{fig_4}(i) exhibits more complex oscillatory dynamics and the simple argument $\arg \langle a \rangle$ does not vary constantly with time as shown in Fig.~\ref{fig_4}(j). 
Thus, $\arg \langle a \rangle$ cannot be considered the asymptotic phase, although it is often used to define power spectra in the analysis of quantum synchronization.
It is noted that the asymptotic phase is quantitatively different from the geometric angle also in the classical limit when the Kerr effect is added. In the strong quantum regime, due to the strong Kerr effect, the difference between the asymptotic phase $\Omega_j$ and simple argument $\arg \langle a \rangle$ can be observed more clearly than in the semiclassical regime or in the classical limit \cite{kato2022definition}. 

\begin{figure} [!t]
	\begin{center}
		\includegraphics[width=1\hsize,clip]{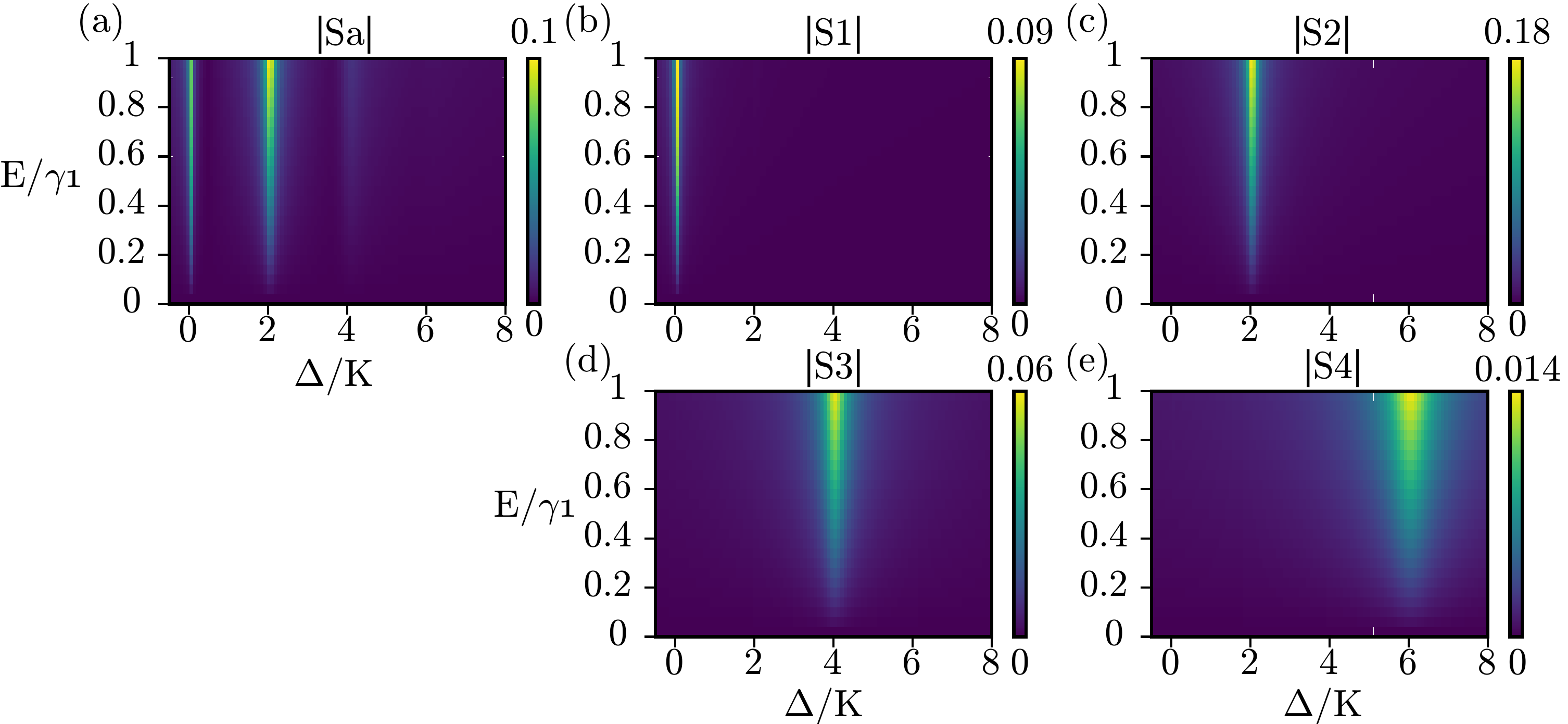}
		\caption{
			Dependence of  the 
			order parameters on the frequency detuning $\Delta$ (divided by $K$) and driving strength $E$ (divided by $\gamma_1$).
			(a) $\abs{S_a}$.
			(b-e) $\abs{S_j}~(j=1,2,3,4)$.
			The parameters are $\gamma_1 = 0.1$ and $(\gamma_{2}, K)/\gamma_{1} = (4, 100)$.
		}
		\label{fig_5}
	\end{center}
\end{figure}

\begin{figure} [!t]
	\begin{center}
		\includegraphics[width=1\hsize,clip]{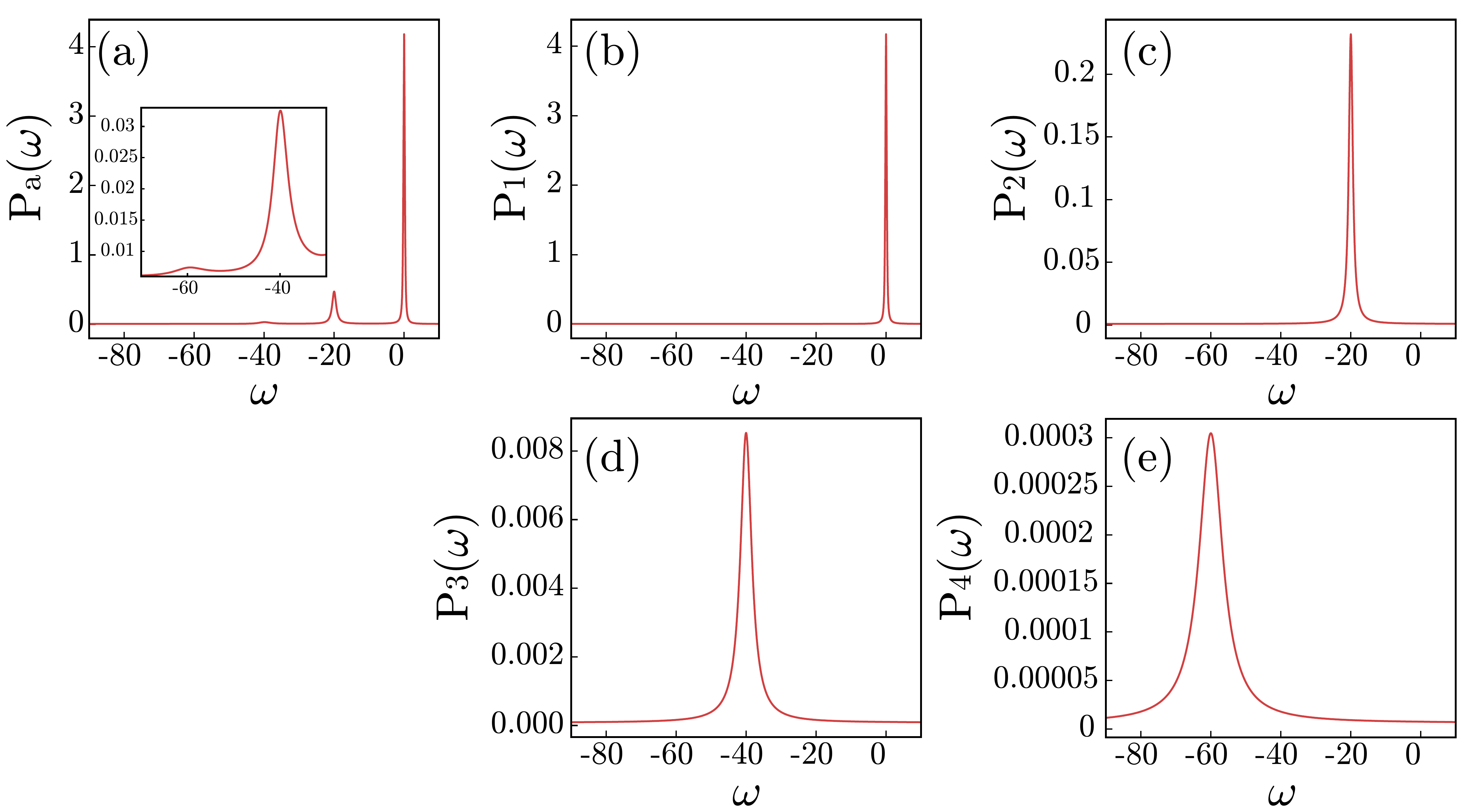}
		\caption{
			Power spectra.
			The parameters are 
			$\gamma_1=0.1$
			and $(\Delta, \gamma_{2}, K, E)/\gamma_{1} = (0, 4, 100, 1)$. 
		(a) $P_a$.
		(b-e) $P_j~(j=1,2,3,4)$. 
		}
		\label{fig_6}
	\end{center}
\end{figure}

\subsection{Revealing multiple phase-locking structure}

We now analyze quantum synchronization of the vdP oscillator with a harmonic drive in the strong quantum regime using the proposed asymptotic phases.

The master equation in the rotating frame of the frequency $\omega_{d}$ 
of the harmonic drive is
\begin{align}
\label{eq:qvdp_me_hd}
\dot{\rho} 
= ( \mathcal{L}_0 + \mathcal{L}_1 ) \rho,
\end{align}
where ${\cal L}_0$ is given by Eq.~(\ref{eq:qvdp_me})
with $H = -\Delta a^{\dag}a + K a^{\dag 2} a^2$
and the harmonic drive is represented by 
$\mathcal{L}_1 \rho= -i \left[ i E (a - a^{\dag}),\rho\right]$,
where
$\Delta = \omega_{d} - \omega_{0}$ is the frequency detuning  
of the harmonic drive from the oscillator
and $E$ is the strength of the harmonic drive~\cite{kato2020semiclassical,lorch2016genuine}. 
We use the same parameters as in Fig.~\ref{fig_1}(a) and vary the detuning parameter $\Delta$
by controlling $\omega_{d}$ while keeping the natural frequency $\omega_0$ fixed.

In Ref.~\cite{lorch2016genuine}, L\"orch {\it et al.} showed that 
this system under strong Kerr effect exhibits multiple phase locking to the harmonic drive at several detuning frequencies $\Delta = 2mK$ ($m=0, 1, 2, \ldots$), i.e., at $\omega_d = \omega_0 + 2m K$, observed as multiple sharp Arnold tongues, while the corresponding classical system exhibits only a single broad Arnold tongue. 
It is stressed that $K$ is the Kerr parameter, hence this is not the ordinary higher-harmonic phase locking at the frequencies $\omega_d = ({p}/{q}) \omega_0$ with $p, q = 1, 2, 3., ...$ ($p/q \neq 1$); the multiple Arnold tongues are related to the periodic transitions between the adjacent discrete energy states.
It is also noted that such multiple phase locking is robust to thermal noise under the finite temperature environment (See Supplemental Material in Ref.~\cite{lorch2016genuine}).

In Ref.~\cite{lorch2016genuine}, the following order parameter $S_a$ and power spectrum $P_a$ defined using the annihilation operator $a$ are used to analyze the system:
\begin{gather}
\label{eq:ord}
S_a = \abs{S_a} e^{i \theta_a} = \frac{\mean{a} }{\sqrt{ \mean{a^{\dag} a}}},
\\
\label{eq:ps}
P_{a}(\omega) = \int_{-\infty}^{\infty} d\tau e^{i\omega \tau}
\left(
\mean{ a^{\dag}( \tau ) a(0)} - \mean{a^{\dag}( \tau )} \mean{ a(0)}
\right),
\end{gather}
where the expectation is taken with respect to the 
steady-state density operator obtained from Eq.~(\ref{eq:qvdp_me_hd}).

Here, in addition to these quantities, we introduce the order parameters and the power spectra in terms of the quantum asymptotic phases (or, more precisely, the corresponding principal Koopman eigenoperators) defined in this study.
They are defined using the left eigenoperators $V_{j}$ of ${\cal L}_0$ as 
\begin{gather}
\label{eq:vord}
S_j  = \abs{S_j} e^{i \theta_j} 
= \frac{\mean{V_j} }{\sqrt{ \mean{V_j^{\dag} V_j}}},
\\
\label{eq:vps}
P_j(\omega) = \int_{-\infty}^{\infty} d\tau e^{i\omega \tau}
\left(
\mean{ V_j^{\dag}( \tau ) V_j(0)} - \mean{V_j^{\dag}( \tau )} \mean{ V_j(0)}
\right),
\end{gather}
where $V_j^{\dag}(\tau) =  e^{{\cal L}_{0}^{*} \tau} V_j^{\dag}(0) = e^{\Lambda_j \tau} V_j^{\dag}(0)$ and $V_j^{\dag}(0) = V_j^{\dag}$ ($j=1,2,3$ and $4$).
Here, $|S_a|$ and $|S_j|$ quantify the phase coherence of the system, while $\theta_a$ and $\theta_j$ characterize the averaged phases of the system relative to the harmonic drive.
We note here that we can use $V_j$, which is defined in the original frame, in the rotating frame with the external drive since the eigenoperators $V_{j}$ changes only its phase factor by the coordinate rotation, i.e.,  $e^{ i \omega_d a^{\dag} a t} V_{j}   e^{ - i \omega_d a^{\dag} a t} = {V}_{j} e^{i \Omega t}$ with a constant $\Omega$ (see Appendix A).
The resulting power spectra in the rotating frame are simply shifted from the power spectra in the original frame by $-\Omega$.
Using these two quantities in Eqs.~(\ref{eq:vord}) and ~(\ref{eq:vps}), we can analyze both the phase coherence and frequency characteristics of the oscillator in quantum synchronization.

Figure~\ref{fig_5} shows the dependence of 
the order parameters $|S_a|$ and $|S_j|$ on the detuning $\Delta$ and strength $E$ of the harmonic drive.
In Fig.~\ref{fig_5}(a) showing $\abs{S_a}$, several Arnold tongues representing phase locking of the oscillator at different frequencies are observed~\cite{lorch2016genuine}.
Figures~\ref{fig_5}(b)-(e) show $\abs{S_j}$ for $j=1, 2, 3$ and $4$, respectively.
It is remarkable that the Arnold tongues in Fig.~\ref{fig_5}(a) are clearly decomposed into individual Arnold tongues around $\Delta = 2(j-1)K$ in Figs.~\ref{fig_5}(b)-\ref{fig_5}(e).
This indicates that the order parameters in terms of the asymptotic phases can capture the phase locking dynamics at 
$\omega_d = \omega_0 + 2(j-1)K$ for different $j$ individually.

Similarly, Fig.~\ref{fig_6} shows the power spectra $P_a$ and $P_j$ for $\Delta = 0$.
Multiple peaks of $P_a$ in Fig.~\ref{fig_6}(a), which indicate multiple phase locking of the oscillator to the harmonic drive, are clearly decomposed into individual peaks around $\omega = \Delta - 2 (j-1) K$  
in Figs.~\ref{fig_6}(b)-(e) for $P_{j}$ ($j=1,2,3$, and $4$).
The Arnold tongue and power spectrum are sharper when the decay rate characterized by $\mbox{Re}\ \Lambda_{j}$ is smaller.
Though not shown, we can also detect even smaller tongues and peaks with $j \geq 5$.

The above results reveal that, in the strong quantum regime, the system behaves like a torus with several fundamental frequencies and each of the associated oscillating mode individually exhibits phase locking to the harmonic drive at the respective frequency~\cite{grindrod2016phase}, resulting in the multiple Arnold tongues and spectral peaks.
Such a torus-like behavior cannot be observed in the semiclassical regime 
where the system behaves like a noisy limit-cycle oscillator with 
a single fundamental frequency, 
because only the principal eigenvalue $\Lambda_1$ is dominant.
In contrast, in the strong quantum regime, due to the effect of quantum noise, visible differences between the imaginary part of the principal eigenvalues $\Lambda_{j}$ ($j=1, 2, 3$, and $4$) arise and also the different branches of eigenvalues become closer to each other, resulting in the torus-like behavior. 

\section{Conclusion}
In this study, we defined multiple quantum asymptotic phases of quantum nonlinear oscillators in terms of the eigenoperators of the adjoint Liouville superoperator associated with several fundamental frequencies from the Koopman operator viewpoint,
which extends our previous definition of the quantum asymptotic phase associated with the slowest decaying mode of the system.
We introduced the order parameters and power spectra in terms of the proposed quantum asymptotic phases and
applied them to the analysis of a quantum van der Pol oscillator with Kerr effect.
We successfully revealed that the multiple phase locking of the system with a harmonic drive at several different frequencies \cite{lorch2016genuine}, which is an explicit quantum signature observed only in the strong quantum regime, can be interpreted as synchronization on a torus rather than on a simple limit cycle.

Though not discussed in this paper, it will also be interesting to introduce the amplitude functions in addition to the quantum asymptotic phases by extending the definition for classical stochastic oscillators \cite{perez2021isostables, kato2021asymptotic}, which can be defined using the eigenfunctions associated with the eigenvalues on the real axis.
The phase-amplitude description that uses both the phases and amplitudes may  be applied to the analysis of quantum complete synchronization \cite{mari2013measures}.
Also, investigating the mutual synchronization between two quantum nonlinear oscillators \cite{lee2014entanglement, walter2015quantum, lorch2017quantum} by using quantum asymptotic phase functions is also a future work. We may be able to introduce a new measure for quantum synchronization between two oscillators, e.g., the normalized correlator \cite{weiss2016noise}, by using quantum asymptotic phase functions.
It may be also interesting to extend our definition of the asymptotic phase for  quantum synchronization in non-Markov open quantum systems \cite{karpat2021synchronization}.

We expect that the proposed definition of the quantum asymptotic phases will serve as a fundamental tool for analyzing strong quantum effects in synchronization~\cite{lorch2016genuine,lorch2017quantum}
and will be useful for future applications of quantum synchronization in the growing fields of quantum technologies.

\section*{Data availability statement }
The data that support the findings of this study are available upon reasonable request from the authors.

\begin{acknowledgements}
Numerical simulations are performed by using QuTiP numerical
toolbox \cite{johansson2012qutip, johansson2013qutip}. We acknowledge JSPS KAKENHI
JPJSBP120202201, JP20J13778, JP22K14274, JP22K11919, JP22H00516, and JST CREST JP-MJCR1913 for financial support. 
\end{acknowledgements}

\appendix
\section{Proof of 
	$e^{ i \omega_d a^{\dag} a t} V_{j}   e^{ - i \omega_d a^{\dag} a t} 
	= {V}_{j} e^{i \Omega t}$}
The time evolution of 
$\tilde{V}_{j}(t) = e^{ i \omega_d a^{\dag} a t} V_{j}   e^{ - i \omega_d a^{\dag} a t}$
is given by 
\begin{align}
	\frac{d \tilde{V}_{j}(t)}{dt} =  i \omega_d a^{\dag} a e^{ i \omega_d a^{\dag} a t} V_{j}   e^{ - i \omega_d a^{\dag} a t} + e^{ i \omega_d a^{\dag} a t} V_{j} (-i \omega_d a^{\dag} a)  e^{ - i \omega_d a^{\dag} a t}
	=  i \omega_d   e^{ i \omega_d a^{\dag} a t}  [a^{\dag} a, V_{j}] e^{ - i \omega_d a^{\dag} a t}.
\end{align}
Because the adjoint superoperator $  \mathcal{L}^{*}_{0}$
couples only the 
elements of the form $ | n + k \rangle  \langle m + k |$ to the elements of the form $ | n + k' \rangle  \langle m + k' |  $ for an arbitrary pair of integers $\{n, m\}$ and integers $k, k'$ 
~\cite{barnett2000spectral, briegel1993quantum},
the eigenoperators $ V_j$  can be explicitly written as 
\begin{align}
V_j =  \sum_{n = 0}^{\infty} v_j(n)| n \rangle \langle n + k |.
\end{align}
Therefore, we obtain $[a^{\dag} a, V_{j}] = - k V_{j}$ and 
\begin{align}
\frac{d \tilde{V}_{j}(t)}{dt} =  i \omega_d   e^{ i \omega_d a^{\dag} a t}  [a^{\dag} a, V_{j}] e^{ - i \omega_d a^{\dag} a t} = -i \omega_d k \tilde{V}_{j}(t).
\end{align}
If we assume $\Omega = -i \omega_d k$, we obtain 
$\tilde{V}_{j}(t) = V_{j} e^{-i \omega_d k t} = V_{j} e^{i \Omega t}$.
%


\end{document}